\documentclass[fleqn,usenatbib]{mnras}

\usepackage{newtxtext,newtxmath}
\usepackage[T1]{fontenc}
\usepackage{ae,aecompl}

\usepackage{graphicx}	% Including figure files
\usepackage{amsmath}	% Advanced maths commands
\usepackage{amssymb}	% Extra maths symbols
\usepackage{xcolor}
\usepackage{enumitem}
\usepackage{placeins}

\newcommand{\Teff}{$T_{\rm eff}$~}  
\newcommand{\logg}{$\log$ g~}
\newcommand{\micro}{$\xi_{t}$~}
\newcommand{\Fe}{[Fe/H]~}
\newcommand{\C}{[C/Fe]~}
\newcommand{\velo}{km s$^{-1}$~}
\newcommand{\Msun}{M$_\odot$~}

\usepackage{soul}
\usepackage{booktabs}

\title[Constraining Nucleosynthesis in 2 CEMP using F.]{Constraining Nucleosynthesis in Two CEMP Progenitors Using Fluorine}

\author[A. Mura-Guzm\'an et al.]{
Aldo Mura-Guzm\'an$^{1,2}$\thanks{E-mail: aldo.muraguzman@anu.edu.au},
D. Yong$^{1,2}$,
C. Abate$^{3}$,
A. Karakas$^{4,2}$,
C. Kobayashi$^{5,2}$,
\newauthor{H. Oh$^{6}$,
{S.-H. Chun$^{6}$},
G. Mace$^{7}$}\\
$^{1}$Research School of Astronomy and Astrophysics, Australian National University, Mount Stromlo Observatory, Cotter Road, Weston Creek, ACT 2611, Australia\\
$^{2}$ARC Centre of Excellence for All Sky Astrophysics in 3 Dimensions (ASTRO 3D), Australia\\
$^{3}$Deep Blue Srl, Piazza Buenos Aires 20, 00198, Rome, Italy\\
$^{4}$School of Physics and Astronomy, Monash University, Clayton, VIC 3800, Australia\\
$^{5}$Centre for Astrophysics Research, University of Hertfordshire, College Lane, Hatfield, AL10 9AB, UK\\
$^{6}$Korea Astronomy and Space Science Institute, 776 Daedeokdae-Ro Yuseong-Gu Daejeon, 34055 Republic of Korea\\
$^{7}$Department of Astronomy and McDonald Observatory, University of Texas at Austin, 2515 Speedway, Stop C1400, Austin, Texas 78712-1205, USA
}

\date{Accepted XXX. Received YYY; in original form ZZZ}

\pubyear{2019}

\begin{document}
\label{firstpage}
\pagerange{\pageref{firstpage}--\pageref{lastpage}}
\maketitle

\begin{abstract}

We present new fluorine abundance estimations in two carbon enhanced metal-poor (CEMP) stars, HE~1429--0551 and HE~1305+0007.  HE~1429--0551 is also enriched in slow neutron-capture process ($s$-process) elements, a CEMP-s, and HE~1305+0007 is enhanced in both, slow and rapid neutron-capture process elements, a CEMP-s/r. The F abundances estimates are derived from the vibration-rotation transition of the HF molecule at 23358.6 \AA~ using high-resolution infrared spectra obtained with the Immersion Grating Infrared Spectrometer (IGRINS) at the 4m-class Lowell Discovery Telescope. Our results include a F abundance measurement in HE~1429--0551 of A(F) = +3.93 ([F/Fe] = +1.90) at [Fe/H] = $-$2.53, and a F upper limit in HE~1305+0007 of A(F) $<$ +3.28 ([F/Fe] $<$ +1.00) at [Fe/H] = $-$2.28. Our new derived F abundance in HE~1429--0551 makes this object the most metal-poor star where F has been detected. We carefully compare these results with literature values and state-of-the-art CEMP-s model predictions including detailed AGB nucleosynthesis and binary evolution. The modelled fluorine abundance for HE~1429--0551 is within reasonable agreement with our observed abundance, although is slightly higher than our observed value. For HE 1429-0551, our findings support the scenario via mass transfer by a primary companion during its thermally-pulsing phase. Our estimated upper limit in HE~1305+0007, along with data from the literature, shows large discrepancies compared with AGB models. The discrepancy is principally due to the simultaneous $s$- and $r$-process element enhancements which the model struggles to reproduce.\\

\end{abstract}


%%%%%%%%%%%%%%%%%%%%%%%%%%%%%%%%%%%%%%%%%%%%%%%%%%

%%%%%%%%%%%%%%%%% BODY OF PAPER %%%%%%%%%%%%%%%%%%

%%%%%%%%%%%%%%%%%%%%%%%%%%%%%%%%%%%%%%%%%%%%%
\section{Introduction}\label{Intro}
%%%%%%%%%%%%%%%%%%%%%%%%%%%%%%%%%%%%%%%%%%%%%

Very Metal-Poor (VMP) stars (\Fe $<$ $-$2.0) observed today are long-lived stars that have witnessed the early beginnings of the Milky Way \citep{Frebel2015}. The chemical compositions in their stellar atmospheres encode information about the first mechanisms of chemical enrichment. That is because VMP stars formed after gas in the Milky Way was enriched by only one or two generations of star formation. Furthermore, many VMP stars obtained their chemical compositions directly through mass transfer from a close companion. For these stars in particular, the stellar atmosphere reveals information about early nucleosynthesis processes in their companion progenitors. Thus, VMP stars are a key to understanding the early evolution of the Galaxy and the first generation of stars. Great efforts have been made by many groups seeking to discover and analyse metal-poor stars \citep[see e.g., review by][]{Beers2005}. An intriguing result from the HK survey by \citet{Beers1985} was the discovery that a considerable fraction of VMP stars exhibit high carbon enrichment. These Carbon Enhanced Metal-Poor (CEMP) stars are defined as those with \C\footnote{In this work we adopted the standard spectroscopic notation $A(\rm{X}) = \log [n(\rm{X})/n(\rm{H})] + 12$ and [X/Y] = $\log[n(\rm{X})/n({Y})]_* - \log[n(\rm{X})/n(\rm{Y})]_{\odot}$ } $>$ 0.7 \citep{Aoki2007} and the number of CEMP stars increases with decreasing metallicities. For example, at [Fe/H] = $-$2.0, the CEMP fraction reaches $\sim$20\%, while at [Fe/H] = $-$4.0, this fraction rises up to $\sim$ 80\% \citep{Placco2014}. In addition to carbon, other element enhancements are found in the stellar atmospheres of these peculiar objects such as light elements and neutron-capture elements. The abundance distribution of these elements in CEMP stars normally follow a few distinctive patterns, which have a direct connection to the enrichment mechanisms that operated in the early Galaxy.\\

Nowadays, CEMP stars have been classified into four subclasses based on the relative abundances of barium and europium, heavy elements produced by the slow ($s$-) and the rapid ($r$-) neutron-capture process, respectively. In this paper we use the classification adopted by \cite{Abate2015a,Abate2015b} \footnote{These definitions varies slightly among other authors \citet[eg.,][]{Beers2005,Aoki2007,Masseron2010}} as described below:

\begin{itemize}
\item \textit{CEMP-s} stars are CEMP stars that satisfy the criteria [Ba/Fe] $>$ +0.5 and [Ba/Eu] $>$ 0;
\item \textit{CEMP-r} stars show abundance enrichments of [Eu/Fe] $>$ +1.0 and [Ba/Eu] $<$ 0;
\item \textit{CEMP-s/r} are CEMP-s stars, [Ba/Fe] $>$ +0.5 and [Ba/Eu] $>$ 0, also enriched with $r$-process elements, i.e. [Eu/Fe] $>$ +1.0;
\item \textit{CEMP-no} stars do not exhibit enhanced abundances in $r-$ and $s$-process elements, i.e. [Ba/Fe] $<$ +0.5.
\end{itemize}

%#######################
%Figure 1
%#######################
\begin{figure*}
		\includegraphics[width=1\linewidth]{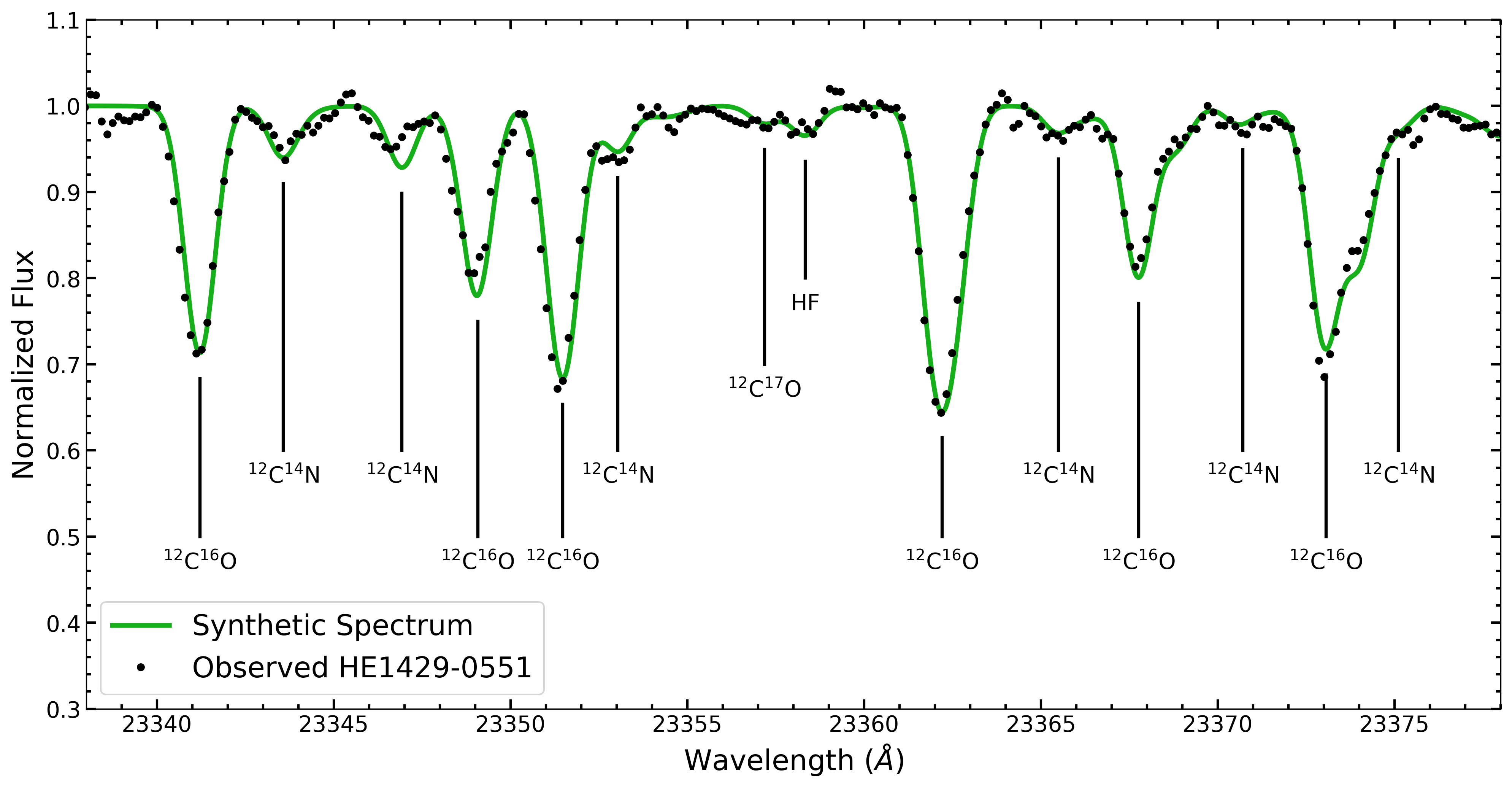}
		\includegraphics[width=1\linewidth]{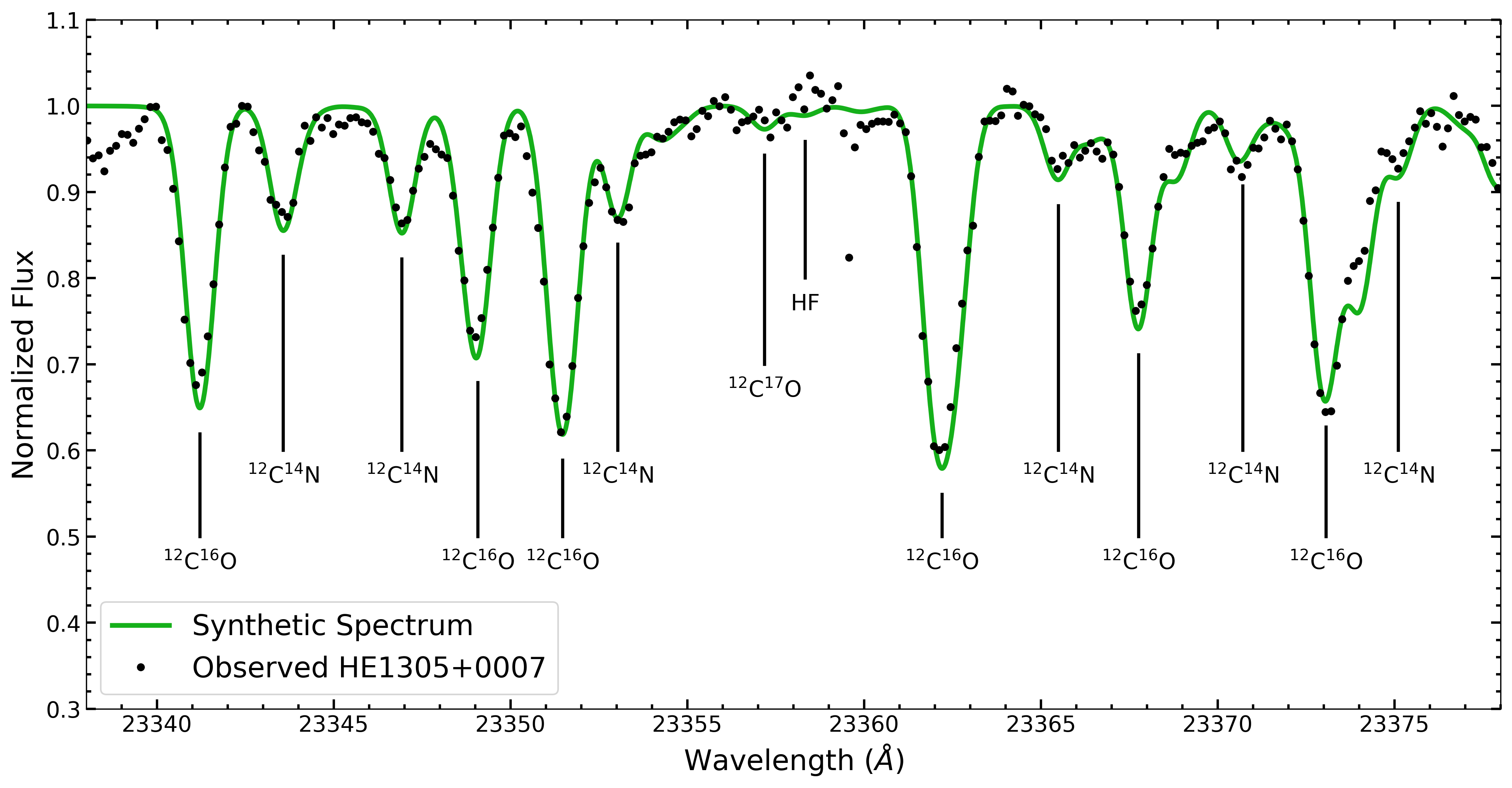}
				\caption{Observed spectra (black circles) and best fitting synthetic spectra (green line) for HE~1428-0551 (upper) and HE~1305+0007 (lower) near the HF line. Note that while the HF line is weak, most of the absorption lines are well fitted. In the lower panel, the HF line region is higher than the continuum suggesting issues with the telluric division (see Section \ref{Results}).}\label{f:synth}
\end{figure*}

\cite{Nomoto2013} summarised the different channels for the formation of CEMP stars. Some of these channels require a first or second generation supernova to pollute the gas cloud that forms the CEMP star with carbon and other metals. Here though, we focus on the mass transfer scenario which takes place in a binary system. According to this scenario the CEMP star receives its carbon and heavy elements from a previous Asymptotic Giant Branch (AGB) companion.\\

AGB stars produce C, N, F and neutron-capture elements in their He-shells, the products of which are mixed to the surface. Strong winds, characteristic of AGBs, expel the chemically enhanced outer layers of the star \citep{Busso1999,Karakas2014}. When the AGB star is in a binary system, some of this wind can be accreted by a low-mass companion star (the present-day CEMP star). As the secondary star slowly evolves, it preserves in its atmosphere the chemical signature of its companion, which at present is an unseen white dwarf \citep{Lucatello2005}.
In a theoretical approach, \cite{Abate2015b}  performed a detailed analysis of AGB nucleosynthesis and binary evolution of some 67 CEMP-s stars from the SAGA database \citep{Suda2008}. Their analysis provides initial parameters and final conditions of the binary system, as well as abundance predictions for several elements in those CEMP-s stars, including fluorine. \\

Fluorine is a particularly rare and fragile element. With just a single stable isotope, $^{19}$F, fluorine is the least abundant of the elements from C to Ca. The synthesis of $^{19}$F requires a series of specific conditions, and once formed, it can be easily destroyed in the stellar interiors \citep{Woosley1988,Jorissen1992,Forestini1992,Mowlavi1996,Meynet2000,
Lugaro2004, Palacios2005,Cristallo2009}. This makes fluorine difficult to produce and extremely sensitive to the environment of its formation site. Three formation sites have been proposed as principal sources of $^{19}$F. AGB stars produce fluorine through complex nuclear reactions starting from $^{14}$N and captures of protons, neutrons and $\alpha$ particles in the reaction $^{14}\rm{N} (\alpha,\gamma) ^{18} \rm{F} (\beta^+) ^{18} \rm{O} (\rm{p},\alpha) ^{15} \rm{N} (\alpha,\gamma) ^{19}\rm{F}$ \citep{Forestini1992}. The F content is then released from the star by stellar winds \citep{Forestini1992,Mowlavi1996,Mowlavi1998,Lugaro2008}. AGB stars are the only observationally confirmed site of F production \citep{Jorissen1992,Werner2005,Zhang2005,Pandey2006}. Other formation sites are theoretically predicted to make fluorine, however, these have not yet been confirmed by observations. Wolf-Rayet stars can also produce fluorine in a similar way to AGB stars. Although production strongly depends on the metallicity and rotation of the star \citep{Meynet2000,Palacios2005,Prantzos2018}. Type II supernovae can synthesize F via neutrino spallation or neutrino process, $\nu$-process \citep{Kobayashi2011b}. High-energy neutrinos $\nu_{\tau}$ ejected during the core-collapse interact with the $^{20}$Ne outer layer producing $^{19}$F \citep{Woosley1988,Woosley1995}.\\

The chemical evolution of fluorine is highly uncertain. Most of the tests for F production are provided by observed F abundances in stars at or near solar metallicities \citep[see e.g.,][]{Jorissen1992,Cunha2008,Abia2010,Recio-Blanco2012,Jonsson2014b,Guerco2019,Ryde2020}. This metallicity bias is due to the difficulties to observe the HF molecular absorption line, the most used absorption feature for the diagnostic of F abundances, in VMP stars. At very low-metallicities, F has been poorly studied with just two reported measurements below [Fe/H] $<$ $-$2: HE~1305+0132 by \citet{Schuler2007}\footnote{This star is in fact more metal rich and a later re-analysis by \cite{Schuler2008} suggested a different result on the observed F abundance (see Section \ref{Comparison_literature} for more details).} and HD5223 by \citet{Lucatello2011} \citep[some other measurements and upper limits are also available in literature, including carbon-normal halo field stars at slightly higher metallicities - e.g.,][]{Lucatello2011,Li2013}. How fluorine is made in the early universe remains unclear. Therefore, studying F in CEMP stars is a direct probe to this question.\\

In this paper, we present an analysis of fluorine abundances in two CEMP stars in order to test nucleosynthesis predictions from \cite{Abate2015b}. Section 2 describes the target selection and observations. Section 3 includes the data reduction and analysis. We present our results and discussion in Section 4. We finalize in Section 5 summarizing our conclusions.

%%%%%%%%%%%%%%%%%%%%%%%%%%%%%%%%%%%%%%%%%%%%%
\section{Target Selection and Observations}
%%%%%%%%%%%%%%%%%%%%%%%%%%%%%%%%%%%%%%%%%%%%%

\subsection{Target Selection}\label{target selection}

%#######################
%Table 1
%#######################
\begin{table*}
\caption{Star ID and stellar parameters from literature data compilation by \citet{Abate2015b} in columns 1-5; our observed fluorine abundance in columns 6-7; modelled abundances from \citet{Abate2015b} in columns 8-9. Additionally, we have included stellar parameters, observed A(F) and [F/Fe] from \citet{Lucatello2011}, and the predicted A(F) from \citet{Abate2015b} for in common stars. Note that F abundance in HE~1305+0007 has been also analysed by \citet{Lucatello2011} (see text).}
\begin{tabular}{l c c c c  c c c}
\toprule
Star ID&				\Teff (K)&	\logg&	\micro (km s$^{-1}$)&	 \Fe& 		A(F)&		[F/Fe]$^{\clubsuit}$&		A(F)$_{Abate}$\\
\hline
HE~1429--0551&		4700&	1.50&	1.38&	-2.53&		3.93&		1.90&		4.40\\
HE~1305+0007&	4655&	1.50& 	1.41&	-2.28&		<3.28&		<1.00&		4.52\\
\midrule
\midrule
\multicolumn{8}{c}{\citet{Lucatello2011} sample}\\
\midrule
CS~22942--019&		5100&		2.5&		-&		-2.5&		<4.2&		<2.1&			5.01\\
CS~29497--034&	4800&		1.8&		-&		-2.9&		<3.5&		<1.8&			-\\
CS~30314--067&		4400&		0.7&		-&		-2.85&		<2.0&		<0.3&			-\\
CS~22948--027&	4800&		1.8&		-&		-2.5&		<3.6&		<1.5&			4.5\\
CS~29502--092&	4890&		1.7&		-&		-3.18&		<3.9&		<2.5&			-\\
CS~30322--023&	4100&		-0.3&	-&		-3.39&		<1.8&		<0.6&			-\\
HE~1152--0355&		4000&		1.0&		-&		-1.27&		3.9&			0.64&			-\\
HE~1305+0007&	4560&		1.0&		-&		-2.5&		<3.5&		<1.4&			4.52\\
HD~187861&			4600&		1.7&		-&		-2.36&		<3.7&		<1.6&			4.61\\
HD~5223&				4500&		1.0&		-&		-2.06&		4.0&			1.44&			3.89\\
HD~122563&			4600&		1.1&		-&		-2.82&		<2.5&		<0.8&			-\\
\midrule
\multicolumn{8}{l}{$^{\clubsuit}$ The adopted solar abundance for F is from \cite{Grevesse2007}, A(F)$_{\odot}$ = 4.56.}
\end{tabular}
\label{t:stellar_params}
\end{table*}

\cite{Abate2015a,Abate2015b} modelled AGB stellar nucleosynthesis and binary evolution at low metallicity and produced a grid of theoretical binary population synthesis models of about $400,\!000$ binary stars with different initial masses and orbital periods. The comparison of this model grid to a sample of 67 CEMP-s stars was used to determine the best-fitting models for the observed chemical abundances, surface gravity, and period (when available) of each individual star of that sample \citep[see][for further details]{Abate2015a,Abate2015b}. We adopted those stellar parameters (Teff, logg, [Fe/H]) and the F abundance predictions of each star from \citet[and references therein]{Abate2015b}, to generate synthetic spectra with the local thermodynamic equilibrium (LTE) stellar line analysis program MOOG \citep[version 2014]{Sneden1973}. Using the vibration-rotation transition of the HF molecule line near 2.3 \micron, we identified potential targets as those showing a synthetic HF absorption line stronger than $\sim$5\% relative to the continuum. From these potential targets, we selected two stars to perform our follow-up observations, HE~1429--0551 and HE~1305+0007.\\

HE~1429--0551 is a CEMP-s star with no reported abundances of Eu. We adopted the stellar parameters compiled from the literature by \cite{Abate2015b}: \Teff = 4700, \logg = 1.50, \Fe = $-$2.53 from \cite{Aoki2007}. A barium abundance of [Ba/Fe] = +1.57 is also reported by \cite{Aoki2007}. Models by \cite{Abate2015b} predicts a F content in HE~1429-0551 of A(F)$_{\rm Abate}$ = +4.40.\\

HE~1305+0007 is classified as a CEMP-s/r object by \cite{Goswami2006} and exhibits large enhancements in $s$- and $r$-process elements. By using high-resolution spectra, \cite{Goswami2006} derived the stellar parameters: \Teff = 4750 K, \logg = 2.0, and \Fe = $-$2.0, as well as a barium abundance of [Ba/Fe] = +2.32, and a remarkable high content in europium, [Eu/Fe] = +1.97. \cite{Beers2007} estimated, using $V-K$ colour and mid-resolution optical and near-IR spectra, the stellar parameters: \Teff = 4560 K, \logg = 1.0, and \Fe = $-$2.5, in addition to a Ba abundance of [Ba/Fe] = +2.9. We adopted the mean values from the stellar parameters, previously mentioned, compiled by \cite{Abate2015b}. This is: \Teff = 4655 K, \logg = 1.5, and \Fe = $-$2.28 (see Table \ref{t:stellar_params}). \cite{Abate2015b} predict a F abundance in HE~1305+0007 of A(F)$_{\rm Abate}$ = 4.52.\\

Both objects appear to be binaries based on previous radial velocity variations. However, orbital solutions have not yet been published. HE~1429--0551 exhibits radial velocity variations above the 3$\sigma$ level, but no clear trend is observed to support orbital motion \citep{Jorissen2016}. HE~1305+0007 radial velocities were first reported by \cite{Goswami2006}, and later by the \cite{Gaia2018}, both studies have uncertainties $>$ 1\velo and a velocity discrepancy $>$ 10\velo. HE~1305+0007 has been flagged as a binary candidate \citep{Arentsen2019}.

\subsection{Observations}

The CEMP stars HE~1429--0551 and HE~1305+0007 were observed using the Immersion Grating Infrared Spectrometer (IGRINS) \citep{Mace2018,IGRINS2,IGRINS1}, mounted on the Lowell Discovery Telescope at the Lowell Observatory. Observations were performed under service mode on 09 and 14 April 2019. The 0.63" slit provided spectral resolution of $R\,\sim$ 45,000, and covered the infrared H and K bands from 1.45 to 2.45 $\mu$m. Although the instrument has a large spectral coverage, for the purpose of this work we use a small fraction of the spectra at  $\sim$2.3$\mu$m.\\

%#######################
%Table 2
%#######################
\begin{table}\caption{Observing Log from Infra-red spectroscopy by IGRINS.}
\begin{scriptsize}
\begin{tabular}{lccccc}
\toprule
Star ID&			RA (2000)&				Dec. (2000)&				UT Date&					Exposure&	$K^{\spadesuit}$\\
&	&	&	&	Time (s)&	\\
\midrule
HE~1429--0551&	14:32:31.56&	-06:04:59.50&		2019 April 09&		8 $\times$ 450&	10.07\\
					 &					  &						 &		2019 April 14&	4 $\times$ 480&	10.07\\
HE~1305+0007&	13:08:04.11&	-00:08:39.60&		2019 April 14&	4 $\times$ 300&	9.60\\
\midrule
\multicolumn{6}{l}{$^{\spadesuit}$ K magnitudes from 2MASS \citep{Cutri2003}}
\end{tabular}
\end{scriptsize}
\label{t:obs}
\end{table}

The entire set of observations were executed under "ABBA" nod sequences. For HE~1305+0007, one ABBA sequence was programmed  with an exposure time for each frame of 300 seconds (i.e., 4 $\times$ 300). HE~1429--0551 was observed twice during the first night and once during the second night with exposure times of 450 and 480 seconds per frame, respectively. The signal-to-noise ratios (S/N), exceed 150 per resolution element in the wavelength-region of interest, at 2.3 \micron, in each star. In addition to the science observations, rapidly rotating hot stars were observed each night, close in time to our program targets, in order to obtain telluric spectra. The telluric spectra are divided into the science spectra of HE~1305+0007 and HE~1429--0551 in order to remove the telluric absorption lines. The program stars and log of observations are listed in Table \ref{t:obs}.

%%%%%%%%%%%%%%%%%%%%%%%%%%%%%%%%%%%%%%%%%%%%%
\section{Data Reduction and Analysis}\label{data reduction}
%%%%%%%%%%%%%%%%%%%%%%%%%%%%%%%%%%%%%%%%%%%%%

Wavelength-calibrated spectra were produced using the IGRINS reduction pipeline package PLP2 \citep{IGRINSpipeline}. PLP2 performs the flat fielding, background removal, order extraction and wavelength calibration. IRAF\footnote{IRAF (Image Reduction and Analysis Facility) is distributed by the National Optical Astronomy Observatory, which is operated by the Association of Universities for Research in Astronomy, Inc., under contract with the National Science Foundation.} tasks were used to combine the individual spectra into a single spectrum and to normalise the continuum. Inspection of the spectrum of HE~1305+0007 (See Figure \ref{f:synth}) indicates that there was likely an issue with telluric removal in the vicinity of the HF line, i.e., the normalized flux exceeds 1.0. We refer to the effect of this last issue on our F abundance estimation in HE~1305+0007 in Section \ref{Results}. \\

The adopted stellar parameters (\Teff, \logg, \Fe) for our sample and its literature source are described in Section \ref{target selection} and Table \ref{t:stellar_params}. The microturbulent velocity \micro, was determined using the relation \micro = $ 4.2 - ( 6 \times 10 ^{-4}$ \Teff $)$ \citep{Melendez2008}.  The abundance dependence due to uncertainties in the adopted stellar parameters for HE~1429--0551 are presented in Table \ref{t:errors}. The reason that we adopted the stellar parameters from the literature instead of derive them directly from the spectra is that 1) Adopting new stellar parameters would lead to inconsistency with the values used in the Abate predictions, and 2) new stellar parameters cannot be easily obtained from the IGRINS spectra alone. This is due to the lack of Fe lines in the wavelength coverage of the instrument used for the estimation of Teff and logg through excitation and ionization equilibrium, respectively. Alternative, photometric temperatures are feasible, but these are C-rich objects and caution should be exercised when using the standard calibrations which are generally based on C-normal objects.\\

%#######################
%Figure 2
%#######################
\begin{figure}
		\includegraphics[width=1\linewidth]{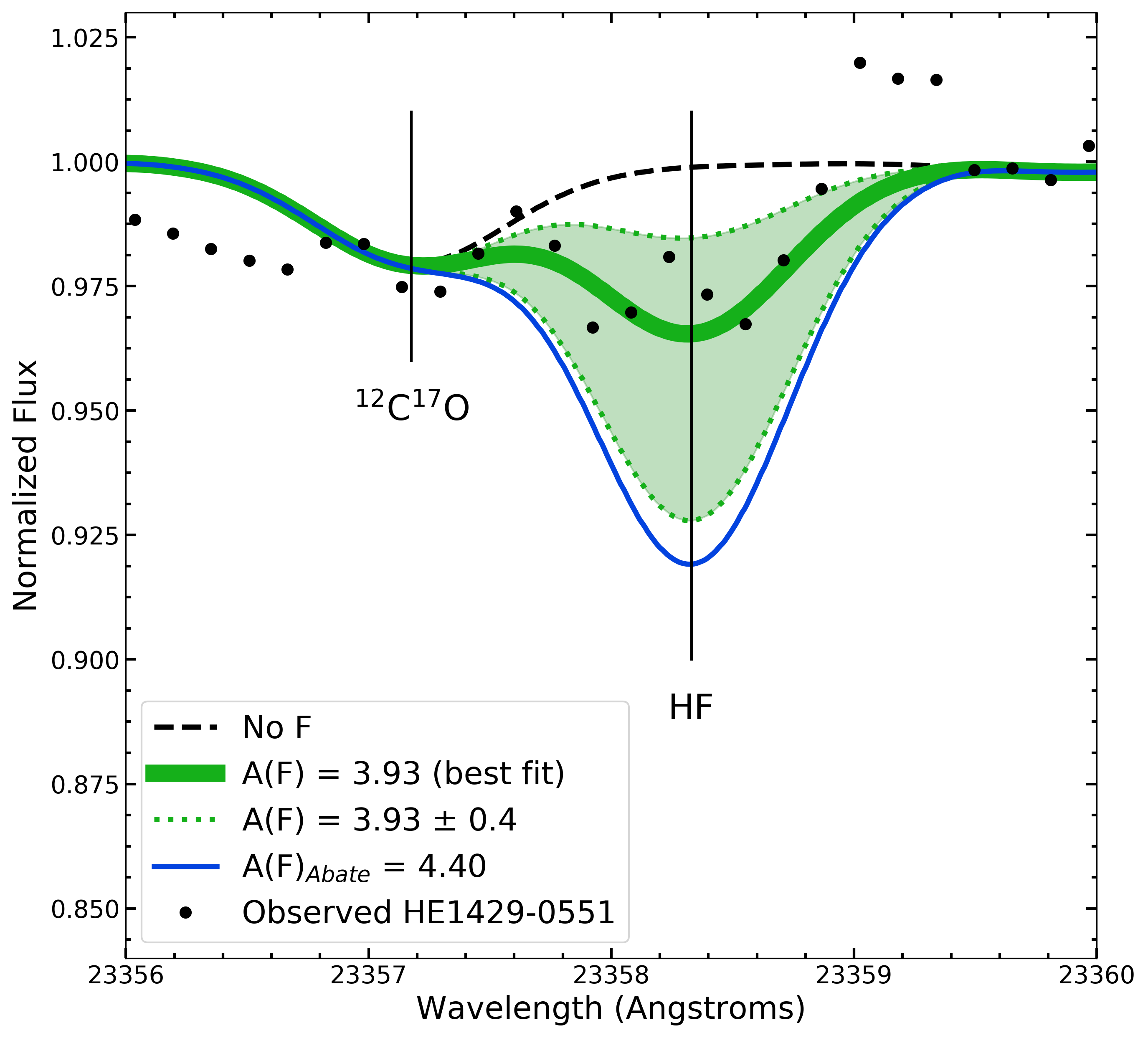}
		\includegraphics[width=1\linewidth]{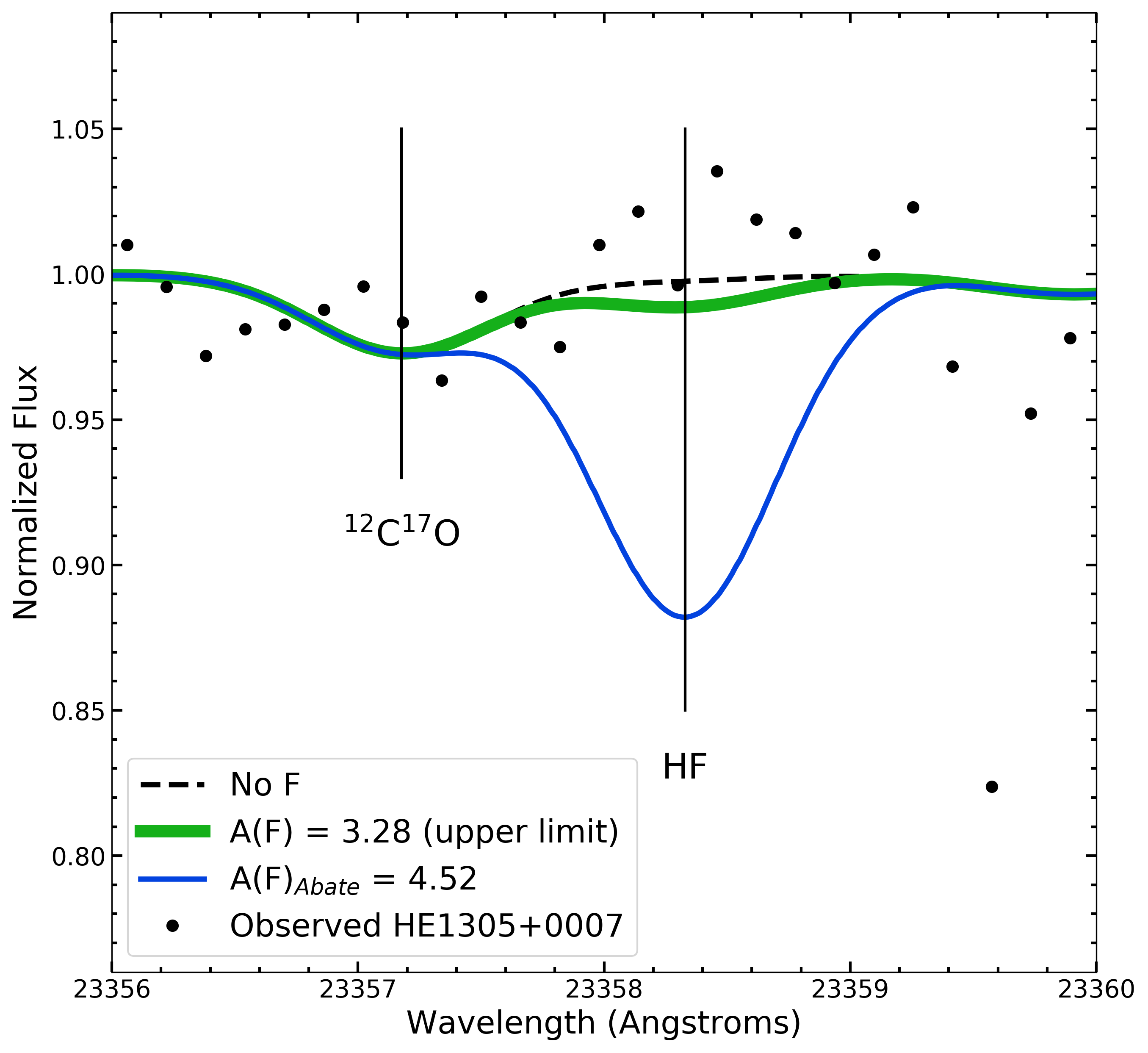}
				\caption{Observed and synthetic spectra near the HF 23358.329\AA\ line for HE~1429--0551 (upper) and HE~1305+0007 (lower). In the upper panel, the solid green line is our best fit for HE~1429--0551 with A(F) = 3.93. The uncertainty region of our fit is shown in shaded green area and delimited by a dotted green line with A(F) = 3.93$\pm$0.4. The predicted F abundance from \citet{Abate2015b} of A(F)$_{\rm{Abate}}$  = 4.40 is shown in the solid blue line. In the lower panel, our upper limit for HE~1305+0007 of A(F) = 3.28 is the solid green line and the predicted value from \citet{Abate2015b} of A(F)$_{\rm{Abate}}$ = 4.52 is the solid blue line.}\label{f:synthzoom}
\end{figure}

%#######################
%Table 3
%#######################
\begin{table}\caption{Fluorine abundance dependence upon model parameter uncertainties for HE~1429--0551.}
\begin{small}
\begin{tabular}{l c c c c c}
\toprule
Star ID&				$\Delta$\Teff (K)&		$\Delta$\logg&		$\Delta$\Fe&	$\Delta$\micro (km s$^{-1}$)&	Total$^{\star}$\\
&							-100&					+0.50&	      			+0.30&				+0.30&					\\
\midrule
HE~1429--0551&		+0.30&			+0.17&			+0.20&			+0.00&			+0.40\\
\bottomrule
\multicolumn{6}{l}{$^{\star}$ The total value is the quadrature sum of the individual abundance}\\
\multicolumn{6}{l}{ dependences.}
\end{tabular}
\end{small}
\label{t:errors}
\end{table}

The fluorine abundances were derived by comparing observed and synthetic spectra. We used the LTE program MOOG \citep{Sneden1973} coupled with one dimensional LTE model atmospheres from the \citet{Kurucz1993} grid to generate the synthetic spectra. Following previous studies of fluorine \citep[eg.,][]{Smith2005}, we studied the vibration-rotation transition of the HF molecule absorption line, HF (1–0) R 9,  at 23358.329 \AA. Although this HF molecular line is not the strongest in the wavelength region, it is free of blends and is considered a reliable indicator of F abundances \citep{Abia2009,Lucatello2011,Jonsson2014a}. We adopted the most recent data for the HF(1–0) R9 molecular line from \citet[and references therein]{Jonsson2014a}; excitation potential $\chi$ = 0.227 and $\log gf$ = -3.962. These HF features were incorporated in our line list and are compatible with the partition function built into MOOG. By doing this we ensure that we are using energies consistent with the partition function integrated in MOOG, avoiding over-estimations in the F abundances previously reported in literature \citep[e.g.,][]{Lucatello2011,Dorazi2013,Jonsson2014b}. We adopted additional molecular data from \citet{Goorvitch1994} and \citet{Sneden2014} for CO and CN, respectively. For the fitting of CO and CN molecular absorption in the spectral region nearby HF, we adopted the observed C abundances from \cite{Aoki2007} for HE~1429--0551, and \cite{Goswami2006} and \cite{Beers2007} for HE~1305+0007, then we adjust O and N in order to fit our observed spectra in these stars. Our best fitting synthetic spectra are shown in Figure \ref{f:synth} and \ref{f:synthzoom}. Note that the $^{12}$C$^{17}$O line at 23358.237 \AA, is slightly blended with the blue wing of the HF line under study. However, this partial blending does not affect significantly the fluorine measurements from the HF(1–0) R9 molecular line \citep{Lucatello2011}.

%%%%%%%%%%%%%%%%%%%%%%%%%%%%%%%%%%%%%%%%%%%%%
\section{Results and Discussion}\label{Results}
%%%%%%%%%%%%%%%%%%%%%%%%%%%%%%%%%%%%%%%%%%%%%

Our derived fluorine abundances are: A(F) = +3.93 ([F/Fe]\footnote{We use the F solar abundance from \cite{Grevesse2007}, A(F)$_{\odot}$ = 4.56, which is also the adopted solar abundance in the comparison literature. For an alternative F solar abundance see \citet{Maiorca2014}.} = +1.90) for HE~1429--0551 and A(F) $<$ +3.28 ([F/Fe] $<$ +1.00) for HE~1305+0007. The adopted metallicity on these CEMP stars are \Fe = $-2.53$ and \Fe = $-2.28$ for HE~1429--0551 HE~1305+0007, respectively.\\

%#######################
%Figure 3
%#######################
\begin{figure*}
		\includegraphics[width=1\linewidth]{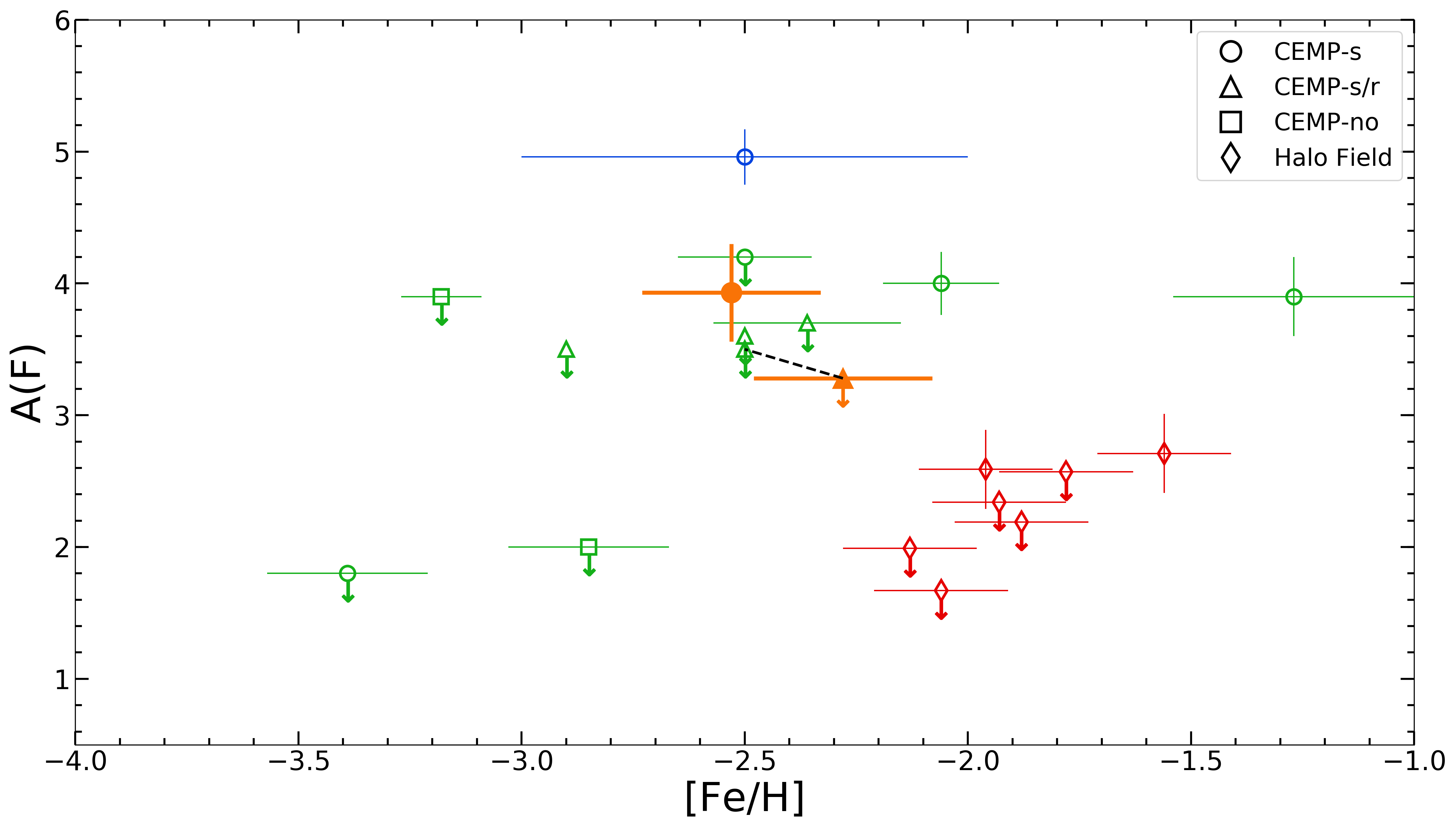}
				\caption{Our fluorine estimations along with literature values for A(F) vs. [Fe/H]. Our fluorine measurement is shown as the filled orange circle for HE~1429--0551 and our upper limit in filled orange triangle for HE~1305+0007. In blue, HE~1305+0132 from \citet{Schuler2007}; in green,  HE~1152--0355,  HD~5223, and other upper limits from \citet{Lucatello2011}; in red, HD~135148, HD~110281, and other upper limits from \citet{Li2013}. The black dashed line shows the differences in the adopted \Fe and the upper limit for the star in common with \citet{Lucatello2011}, HE~1305+0007.}\label{f:F vs Fe}
\end{figure*}

\begin{figure}
		\includegraphics[width=1\linewidth]{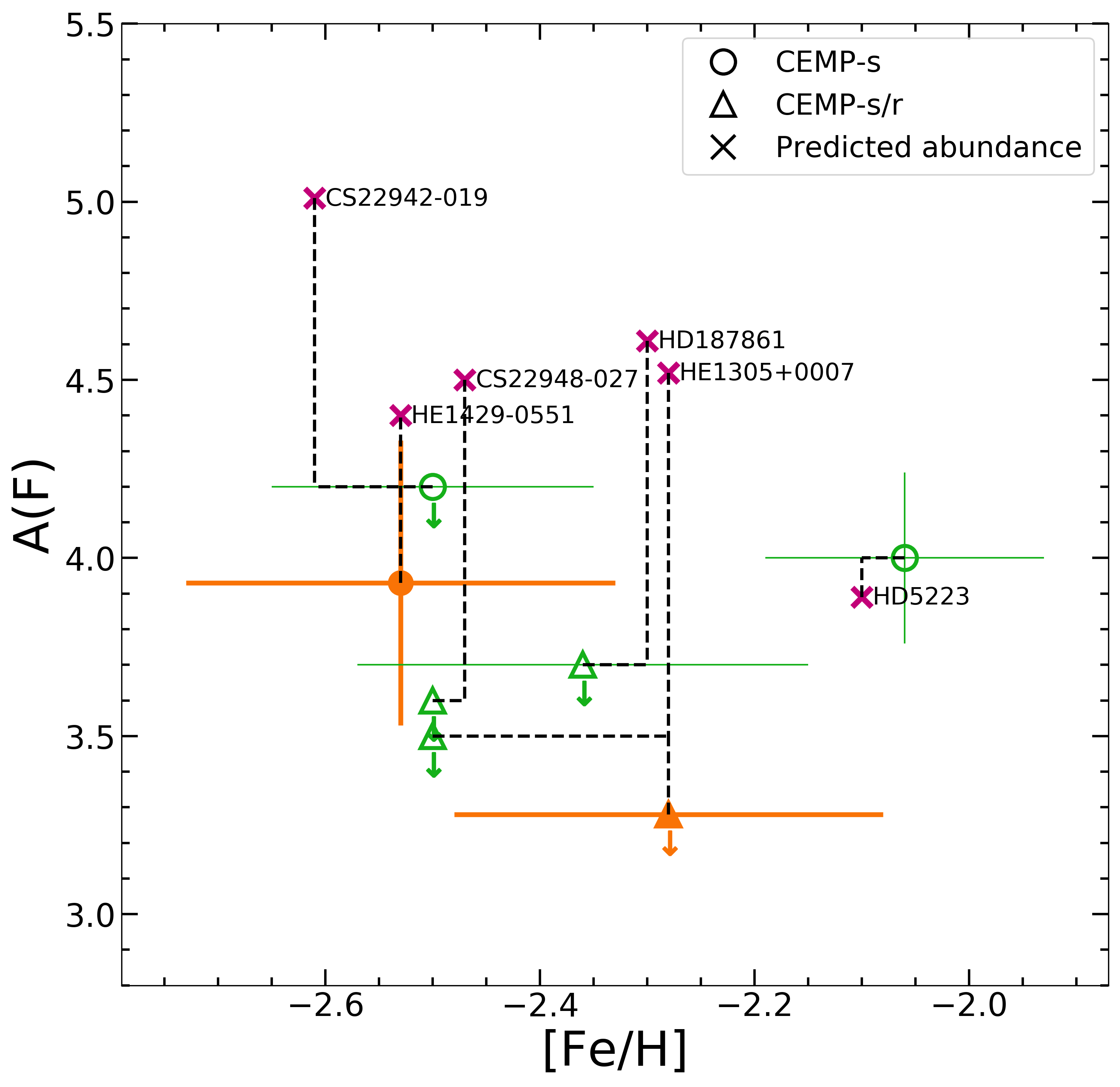}
				\caption{A(F) versus [Fe/H] for our data (orange symbols) and \citet{Lucatello2011} (green symbols). The \citet{Abate2015b} predictions are presented in magenta x symbols. For each star, the observed and theoretical abundances are connected by a black dashed line. Note the differences in the adopted [Fe/H] between \citet{Abate2015b} and \citet{Lucatello2011} (see text for details). HE~13005+0007 is in common between our sample and \citet{Lucatello2011}, and our adopted [Fe/H] for this star also differs from their work. The fluorine abundance measured in HD~5223 by \citet{Lucatello2011} are well reproduced by the theoretical model. All the upper limits are substantially overestimated by \citet{Abate2015b}, including HE~1305+0007.}
				\label{f:Carlo_Lucatello}
\end{figure}

In Figure \ref{f:F vs Fe} we plot our derived A(F) abundances against \Fe along with the literature data of F abundances on CEMP stars -i.e., \cite{Schuler2007,Lucatello2011}. In this figure, we also include literature values of F abundances in carbon-normal halo field stars from \cite{Li2013} as a comparison. Figure \ref{f:Carlo_Lucatello} compares our fluorine abundance estimations with those predicted by \citet{Abate2015b}. In Figure \ref{f:Carlo_Lucatello}, additionally to our results, we include part of the \cite{Lucatello2011} sample which is in common with \citet{Abate2015b}. For clarity, we connect the observed F abundances with the predictions using a black dashed line. The horizontal and vertical sections of the connecting line corresponds to differences in the adopted metallicities and F abundances estimates, respectively, by \cite{Lucatello2011} and \citet{Abate2015b}. We show in Figure \ref{f:synth} and \ref{f:synthzoom} our best synthetic fit for our F measurement in HE~1429--0551 and F upper limit in HE~1305+0007. In addition to our best fit, we include in Figure \ref{f:synthzoom} a synthetic HF absorption line using the predicted F abundance by \citet{Abate2015b} for comparison. We suspect that the telluric absorption in our data for HE1305+0007 has not been fully corrected since the normalized flux in the HF line is higher than 1.0. However, the extent of this effect in our derived upper limit is expected to be small. The predicted strength of the HF line at very low-metallicities is weak, which is the principal reason for the high S/N ratio in our observations. We note that there is a $\sim3\%$ excess in the normalized flux of HE~1305+0007 (see lower panel in Figure \ref{f:synthzoom}). If there was an HF feature, we assume that it would have been detected if it were stronger than $\sim5\%$ relative to the continuum. Moreover, the typical error in the observations of F abundances is $\sim$0.4 dex. This error value would account for this telluric effect if the line were observed, without substantially changing our results. Naturally, more observations in HE~1305+0007 will help to clarify the F abundance in this star. For the case of HE~1429--0551, the telluric contamination was sufficiently corrected. Despite the weakness of the line (3.5\% relative to the continuum), a reduced $\chi^2$ value of 2.2 in the region of interest supports our F measurement in HE~1429-0551.\\

The study of F involves several sources of uncertainty. We introduce the general uncertainties in both the nucleosynthesis and observations as follows. The production of the $^{19}$F isotope in AGB stars takes place in the He intershell layer through complex nuclear reactions. Once formed, fluorine is dredged up to the surface during the third dredge up (TDU) along with C and $s$-process elements \citep{Forestini1992,Mowlavi1996,Mowlavi1998,Lugaro2004}. As mentioned before, the principal reaction pathway for the production of $^{19}$F in AGB stars is $^{14}\rm{N} (\alpha,\gamma) ^{18} \rm{F} (\beta^+) ^{18} \rm{O} (\rm{p},\alpha) ^{15} \rm{N} (\alpha,\gamma) ^{19}\rm{F}$ \citep{Forestini1992}. The protons needed in the reaction are supplied via $^{14}\rm{N}(\rm{n,p})^{14}\rm{C}$, which in turn is provided with neutrons by the $^{13}\rm{C}(\alpha,\rm{n})^{16}\rm{O}$ reaction. This last neutron source, $^{13}\rm{C}(\alpha,\rm{n})^{16}\rm{O}$, also provides neutrons for the $s$-process and its efficiency depends on the amount of $^{13}\rm{C}$ present in the He intershell. Some $^{13}\rm{C}$ existent in the intershell comes from CNO cycling in previous interpulse phases, although it is not enough to achieve the AGB $s$-process element enhancements. As an additional source of $^{13}\rm{C}$, some protons can be partially mixed into the intershell producing $^{13}\rm{C}$ from the abundant $^{12}\rm{C}$ in the region through the reaction $ ^{12}\rm{C} (\rm{p,}\gamma) ^{13}\rm{N}(\beta^+\nu)^{13}\rm{C}$ \citep{Iben1982b}. This reaction generates a thin layer abundant in $^{13}\rm{C}$ and $^{14}\rm{N}$ at the top of the intershell, namely the '$^{13}\rm{C}$ pocket' \citep{Iben1982a}, which enrichment will depend on the amount of protons injected into the intershell, the partial mixing zone (PMZ). The production of F and its relation to the '$^{13}\rm{C}$ pocket', suggest a positive correlation between abundances of F and $s$-process elements at the surface of the AGB star (see Section \ref{Ba-F}). However, some nuclear reaction rates are still not well known. For example, the $^{19}$F($\alpha , p$)$^{22}$Ne reaction is responsible for the destruction of $^{19}$F. Although this reaction is one of the most important parameters for a proper theoretical analysis of F nucleosynthesis in AGB stars, it is also one of the principal uncertainties in the nuclear reaction rates \citep{Lugaro2004}. Other sources of uncertainty in the nucleosynthesis of $^{19}$F include the shape of H profile of the partial mixed zone, convective mixing and mass loss.\\

From an observational perspective, F abundances are difficult to measure. The formation of the HF molecule is very sensitive to \Teff in the atmosphere and is the main source of uncertainty in the adopted stellar parameters \citep{Jorissen1992,Jonsson2014a,Jonsson2014b,Lucatello2011}. The metallicity also affects the stellar structure in the outer layers of the star, changing its temperature profile and thus the HF absorption feature \citep{Lucatello2011}. Additionally, telluric corrections are necessary due to the high contamination in the infrared, generally blending the HF line. 3D and/or non-LTE (NLTE) effects can also play a role in the analysis of observed F abundances. Currently, 3D corrections for fluorine abundances have been explored by \cite{Li2013}, but further investigation might be necessary. NLTE corrections are not available yet for the HF molecule, and the strength of its effect remains unknown. These observing issues makes fluorine hard to detect. At low metallicities, previous to this work, 11 estimations of fluorine abundances have been reported in CEMP stars \citep{Schuler2007,Lucatello2011}: three measurements and eight upper limits. Carbon-normal halo field stars have been observed as well in a sample of seven stars, including two measurements and five upper limits \citep{Li2013}. With an observational success of $\sim28\%$ measurements, the small observational data of F measurements in CEMP stars is not enough to be properly compared with F production in low-metallicity AGB nucleosynthesis models. However, the very presence of F (and enrichment), simultaneously with C and the characteristic $s$-process elements distribution in CEMP-s stars, indicate very strong evidence for AGB stellar nucleosynthesis.

%#######################
%Figure 4
%#######################

\subsection{Comparison with Literature Data}\label{Comparison_literature}

\cite{Schuler2007} measured F for the first time in a CEMP(-s) star, HE~1305+0132. This was the most metal-poor star where a F detection has been reported with A(F) = +4.96 ([F/Fe] = +2.90) at \Fe = $-$2.5, the highest by over $\sim$ 1 dex from all other literature data at very low-metallicities. \cite{Lugaro2008} showed that the observed C and F abundances by \cite{Schuler2007} in HE~1305+0132 can be justified by the binary mass transfer scenario of CEMP-s stars. Their model was able to reproduce the abundances by considering: 1) accretion of $3-11 \%$ of the mass loss from the companion star during the AGB phase; 2) dredged-up intershell material, no less than $\sim$ 0.2 M$_\odot$, into the convective envelope by TDU. However, the analysis by \cite{Schuler2007} used older HF molecular transition data from \cite{Jorissen1992} with an excitation potential $\chi = 0.48$ eV. This value is not compatible with the partition function in the MOOG program \citep{Sneden1973}, leading to an offset in the reported F abundance of $\sim +0.3$ dex. \citep[See e.g.,][]{Lucatello2011,Jonsson2014a}. Moreover, the adopted metallicity for HE~1305+0132 (\Fe = $-$2.5 $\pm$0.5) by \cite{Schuler2007} was derived by \cite{Beers2007} using mid-resolution optical and near-IR spectra. Later, \cite{Schuler2008} reported a revised, and higher, metallicity of \Fe = $-$1.92 dex for HE~1305+0132 using high-resolution optical spectra, $0.6$ dex higher than previously estimated. A more metal-rich model atmosphere contains outer layers cooler than the previous metal-poor model atmosphere used in \cite{Schuler2007}. This metallicity difference can induce a lower F abundance in the measurement of HE~1305+0132 by \cite{Schuler2007}, comparable to those [F/Fe] abundances found in other CEMP objects and the new values obtained in this paper. Additionally, our new F measurement in HE~1429--0551 makes this object the most metal-poor star where F has been detected. (Upper limits have been obtained at similar metallicity \citep{Lucatello2011}).\\

\cite{Lucatello2011} presented fluorine measurements and upper limits in 10 CEMP stars. Their sample included 8 CEMP-s and 2 CEMP-no stars. Fluorine was measured in 2 CEMP-s stars. For the rest of the sample, just upper limits are reported, including the 2 CEMP-no stars (See figure \ref{f:F vs Fe}). HE~1305+0007 is in common between \cite{Lucatello2011} and our sample. In both studies, an upper limit was derived for this star. Our upper limit is 0.22 dex lower than the one obtained by \cite{Lucatello2011} (A(F) $<$ 3.5, [F/Fe] $<$ 1.4). This discrepancy could be due to the differences in the adopted stellar parameters, and the different stellar atmosphere models used in the analysis of HE~1305+0007. We adopt the compiled stellar parameters from \cite{Abate2015b} (See Section \ref{target selection}, and Table \ref{t:stellar_params}), while \cite{Lucatello2011} adopted stellar parameters from \cite{Beers2007}: \Teff$~$ = 4560, \logg$~$ = 1.0 and \Fe = $-$2.5. A re-analysis of our F detection in HE~1305+0007 using the stellar parameters adopted by \cite{Lucatello2011} generates an upper limit of A(F) $<$ 3.11, [F/Fe] $<$ 1.05. This differs from our original analysis  by  $\Delta$A(F) = $+$0.17 and $\Delta$[F/Fe] = $-$0.05 dex. This is principally due to the high sensitivity of the adopted absorption HF line to temperature and its moderate sensitivity to metallicity. The differences in the model atmosphere used in both analyses also have an effect in the F measurements. \cite{Lucatello2011} used spherical C-enhanced models which contribute to slightly higher F abundances for HE~1305+0007 in their study  \citep[see Section 3]{Lucatello2011}. Additionally, issues in the removal of telluric features near the HF line described before, could also affect our derived F upper limit in HE~1305+0007. However, our upper limit is comparable with that upper limit derived by \citet{Lucatello2011} in this star.\\

More recently,  \cite{Li2013} derived F abundances and upper limits in 7 carbon-normal metal-poor giant stars in the halo: two measurements and five upper limits (See Figure \ref{f:F vs Fe}). Their measurements correspond to HD110281 with A(F) = 2.71 at \Fe = $-1.56$, and HD135148 with A(F) = 2.59 at \Fe = $-$1.96. HD135148 has the highest C abundance in their sample and is a spectroscopic binary with a companion star with the approximate mass of a white dwarf \citep[e.g.,][]{Carney2003}. \cite{Li2013} also applied 3D abundance corrections on the observed sample using 3D models from the CIFIST atmosphere grid \citep{Ludwig2009}. Since the grid of CIFIST do not have 3D models at low gravities as their star sample, four 3D models with comparable stellar parameters but higher gravity were selected to inspect the trend of the abundance corrections with changing \Teff and \logg. \cite{Li2013} suggested minor 3D effects for the low-gravity giants in their sample and substantial corrections (> 0.1 dex) for gravities \logg $>$ 2.0, at \Fe = $-$2.0.

\subsection{Observations v/s. Theoretical Models}

The F abundance in CEMP-s stars is assumed to come from an AGB companion (now an unseen white dwarf) which also synthesized C and $s$-process elements \citep{Lugaro2004,Ivans2005,Lucatello2005,Karakas2007,Thompson2008,
Lugaro2008}. Within this picture, theoretical models have to consider not just AGB stellar nucleosynthesis at low-metallicities, but also the binary evolution. \cite{Abate2015a,Abate2015b} have undertaken an appropriate theoretical study (see Section \ref{target selection}). The initial abundances used in \cite{Abate2015a,Abate2015b}, including fluorine, were adopted from the Galactic Chemical Evolution (CGE) model by \citet{Kobayashi2011a}. In this CGE model, the contribution from massive stars is set to minimum and does not include that from WR stars. Although \cite{Kobayashi2011b} have suggested that the F production through neutrino-process can be increased by a factor of 100, this is negligible once mass transfer occurs from AGB stars in the initial-mass range [0.9,2.5] \Msun.\\

Stars HE~1429--0551 and HE~1305+0007 were originally included in the \citet{Abate2015b} sample and we can therefore compare our newly-determined fluorine abundances with the predictions.\\

\textit{HE~1429--0551.} Our observed F abundance, A(F) = +3.93, is slightly overpredicted but in reasonable agreement with the theoretical predictions from \cite{Abate2015b}, A(F)$_{\rm Abate}$ = +4.40. In addition to our F measurement, observed abundances of C and neutron-capture elements by \cite{Aoki2007} are well reproduced by the model for HE~1429--0551. Therefore, the observed chemical pattern of HE~1429--0551 supports the formation scenario on a binary system through mass transfer from a former AGB companion, even though its binary status has not been observationally confirmed as noted in Section \ref{target selection}. The best-fit model from \citet{Abate2015b} for HE~1429--0551 correspond to a very long-period binary with an initial period $P_i = 1.37 \times 10^5$ days. The initial masses of the primary star (the CEMP progenitor) and secondary star (the CEMP star) are $M_{1,i} = 1.30$ \Msun and $M_{2,i} = 0.71$ \Msun, respectively. The adopted PMZ mass of the primary star is $M_{\rm{PMZ}} = 10^{-3}$ \Msun and the calculated accreted mass by the secondary stars is $M_{\rm{acc}} = 0.17$ \Msun. The current estimated period of the system is $P_f = 2.3 \times 10^5$ days.\\

%#######################
%Figure 5
%#######################
\begin{figure*}
		\includegraphics[width=1\linewidth]{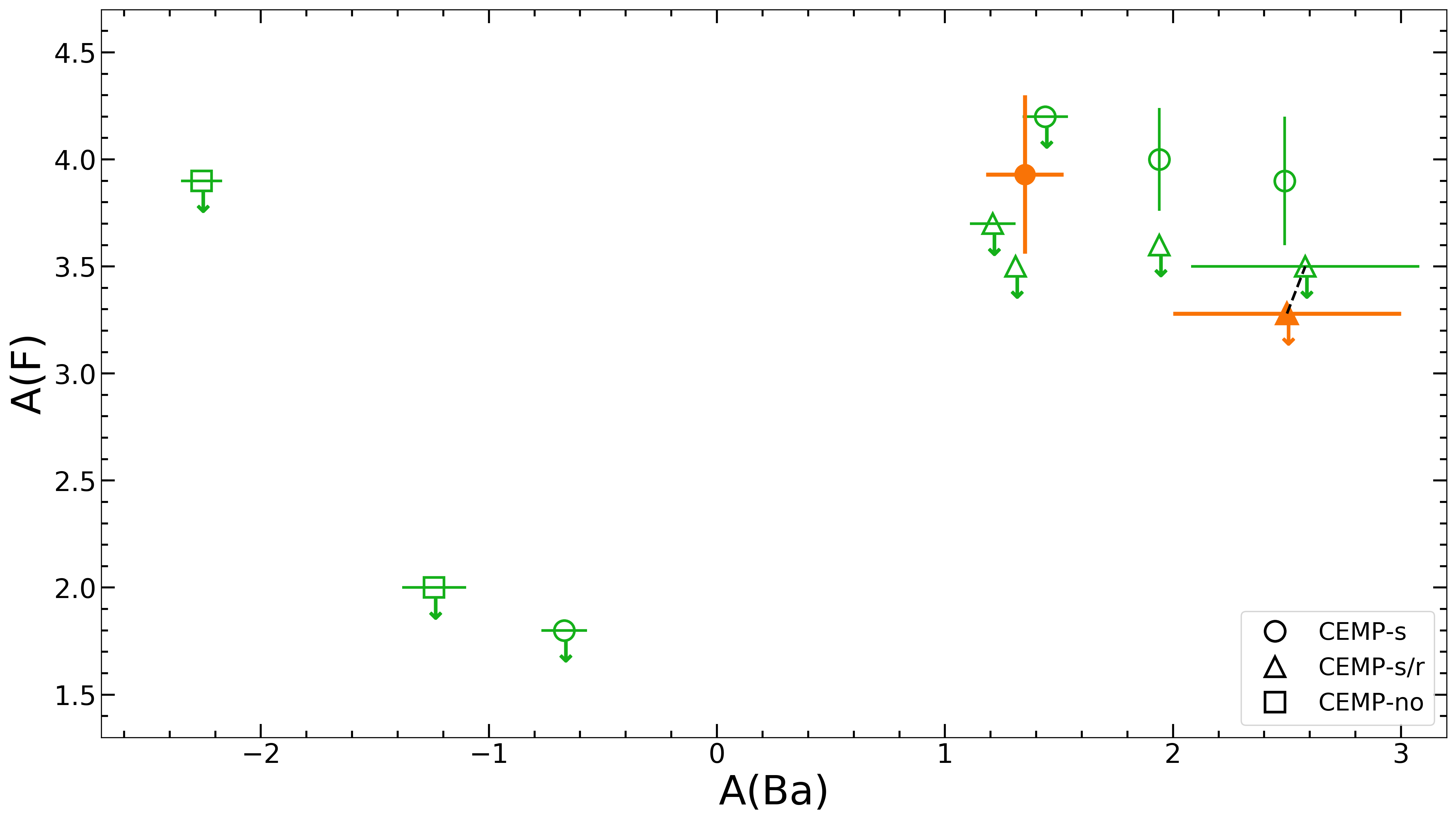}
				\caption{A(F) and A(Ba) abundances for CEMP-s, CEMP-s/r and CEMP-no objects from \citet{Lucatello2011} and our sample. Symbols are the same as Figure \ref{f:F vs Fe} (carbon-normal field halo stars are excluded). The black dashed line shows the differences in the adopted A(Ba) and the estimate upper limit in our in common observed star with \citet{Lucatello2011}, HE~1305+0007.}\label{f:Ba-F}
\end{figure*}

\textit{HE~1305+0007.} Our F abundance upper limit, A(F) $<$ +3.28, does not agree with the theoretical prediction from \cite{Abate2015b}, A(F)$_{\rm Abate}$ = +4.52 dex. This is 1.24 dex higher than our estimated upper limit in this star (see Figure \ref{f:synthzoom} and \ref{f:Carlo_Lucatello}). Ours and the \cite{Lucatello2011} upper limit in this star, both disagree with the predicted abundances. The disagreement is very likely due to the fact that HE~1305+0007 is classed as a CEMP-s/r star, which is a separate class of CEMP stars whose origins may be completely different from CEMP-s \citep[the origin of CEMP r/s from an AGB nucleosynthesis point of view was also discussed by][]{Lugaro2012}. Our observation, as well as literature data from \citet{Lucatello2011}, suggest a lower fluorine production by the HE~1305+0007 progenitor that cannot be reproduced by the nucleosynthesis model adopted by \cite{Abate2015b} while simultaneously reproducing the very enhanced abundances of the neutron-capture elements. As discussed by  \citet{Abate2015b}, in most CEMP-s/r stars the abundances of heavy neutron-capture elements (e.g. Ba, La, Ce, Pb, but also Nd, Sm, and Eu) are strongly enriched and to reproduce their enhancements, the algorithm determining the best-fitting model selects a solution where a large mass of very enriched material is transferred from the AGB primary star to the main sequence secondary star of the binary system. Consequently, the best-fitting model tends to overestimate the abundances of the light elements (e.g. C, N, F, Na, Mg), which are typically enhanced to a similar extent as "canonical'' CEMP-s stars. This result was interpreted as an indication that CEMP-s/r star might have undergone a different nucleosynthesis history compared to CEMP-s stars, the abundances of which are generally well-reproduced by AGB nucleosynthesis models \citep{Lugaro2012,Abate2015b}.

\citet{Hampel2016,Hampel2019} used a one-zone nucleosynthesis model with neutron-captures operating at neutron densities intermediate to the $s$- and $r$-process, the $i$-process. Their adopted temperatures are consistent with the He-shell temperatures from a previous work from \cite{Stancliffe2011}. The modelled abundances by \citet{Hampel2016,Hampel2019} can satisfactorily reproduce the observed abundances of C and $s$-process elements, as well as the $r$-process elements present in HE~1305+0007. Currently, F abundances from $i$-process models are not available in the literature that we are aware of. It would be very interesting to examine and test if those models can also reproduce the observed abundances of fluorine as well as the $s/r$ (or $i$) chemical pattern.\\

Five stars in the sample from \cite{Lucatello2011} are in common with the theoretical work of \cite{Abate2015b}, including HE~1305+0007 which is also in common with this work. Figure \ref{f:Carlo_Lucatello} shows the comparison between the predictions from \cite{Abate2015b} with the observations from \cite{Lucatello2011} and ours. There are some differences in the stellar parameters adopted by \cite{Lucatello2011} and \cite{Abate2015b} (i.e. \Teff and \Fe). The differences in the adopted \Fe and the predicted-observed abundances for each star are shown in Figure \ref{f:Carlo_Lucatello} using a connecting black dashed line.

The fluorine abundances observed by \cite{Lucatello2011} include CEMP-s and CEMP-s/r. Their measurement in the CEMP-s HD~5223 is in good agreement with the model. However, all their upper limits are highly overestimated by the model predictions from \citet{Abate2015b} (see Figure \ref{f:Carlo_Lucatello}). The inconsistencies with the upper limits of the CEMP-s/r stars CS~22948--027, HD~187861 and HE~1305+0007 can be accounted for by the issues of the model on reproducing the s/r pattern in addition to the differences in the adopted stellar parameters mentioned before. Nevertheless, our upper limit in HE~1305+0007 was derived using the same stellar parameters as in \cite{Abate2015b}, and we found similar discrepancy of 1.24 dex. Additionally, the upper limit in the CEMP-s star CS~22942--019 estimated by \cite{Lucatello2011} is highly overestimated by the model. The inconsistency with the model in this star cannot be justified by just the differences in the adopted stellar parameters.

\subsection{Ba and F abundances in CEMP stars}\label{Ba-F}

In CEMP-s stars, abundances from s-process elements are expected to correlate with those of F given that both are produced in AGB stars \citep{Lugaro2004}, the CEMP-s progenitors. However, this might not always be the case since the abundance of Ba in the envelope reaches an equilibrium value when the $s$-process operates at low metallicity \citep{Lugaro2012}. In contrast, F does not have an equilibrium value and its nucleosynthesis is more sensitive to the thermodynamic conditions in the intershell of the AGB star \citep{Mowlavi1996,Mowlavi1998, Lugaro2004}. Progenitors of CEMP-s/r stars have to account for nucleosynthesis mechanisms able to enhance both $s$- and $r$-process elements, as well as C . Although the $i$-process can explain C and neutron-capture elements in the CEMP-s/r chemical pattern \citep[e.g.,][]{Hampel2016,Hampel2019}, the effect of this process on the F production is unknown. The origin of the chemical pattern observed in CEMP-no objects still remains unclear. A variety of possible formation scenarios are proposed in literature \citep[e.g., ][]{Umeda2005,Masseron2010,Meynet2010,Chiappini2013,Nomoto2013,Limongi2018}. Measured abundances of Fluorine (or strong upper limits) in CEMP-no stars would give useful information on the chemical origins of CEMP-no stars.\\

We show in Figure \ref{f:Ba-F} the absolute abundances of Ba and F from \cite{Lucatello2011} and our sample, in green and orange, respectively. These two sample are the only data available for F and Ba in CEMP stars. Note that Ba abundances cannot be derived from the infrared K-band due to the lack of features in this wavelength region. Hence, the Ba abundances shown in Figure \ref{f:Ba-F} are adopted from the literature. For our sample, we adopt the Ba abundances from the compilation by \cite{Abate2015b} -- i.e., \cite{Goswami2006} and \cite{Beers2007} for HE~1305+0007, and \cite{Aoki2007} for HE~1429--0551. The details for Lucatello's adopted Ba abundances can be found in \citet[][Table 2]{Lucatello2011}. In addition to the inhomogeneities introduced by the Ba abundances in Figure \ref{f:Ba-F}, the estimated fluorine abundances are mostly upper limits. These limitations make it difficult to quantify any trend if present. However, the CEMP-s upper limit at A(Ba) $\sim$ $-0.6$ and the CEMP-s measurements at higher Ba abundances suggest a possible Ba-F correlation in CEMP-s stars. We also note that the 3 CEMP-s stars with measured F abundances at A(Ba) $>$ 1 are higher than those upper limits from CEMP-s/r stars. This suggests that while the Ba ($s$-process elements proxy) production in CEMP-s and CEMP-s/r progenitors is indistinguishable, the F production might be more efficient in CEMP-s than in CEMP-s/r progenitors. Unfortunately, due to the lack of measurements and the large spread in the only two F upper limits in CEMP-no stars, no valuable information can be extracted from its F estimates.\\

Very recently, \citet{Ryde2020} compared the current literature data on F abundances at higher metallicities, in addition to their own sample, with O and the $s$-process element Ce. They showed a flat trend when studying the [F/Ce] ratio as a function of \Fe for intermediate metallicities, $-$0.6 $<$ \Fe $<$ 0.0. This result strengthens their conclusion on AGB stars as the dominant source of F in that metallicity range. However, their sample includes two stars below \Fe $<$ $-0.8$ exhibiting large s-process element (Zr, La and Ce) enhancements yet surprisingly low F abundances. \citep[see Section 5.6 and Figure 10 from][]{Ryde2020}. An inspection of radial velocities in these two stars by \citet{Ryde2020} did not show any signs of binarity, weakening the possibility of pollution from an AGB companion. A similar result was found by \citet{Cunha2003} in $\omega$ Cen stars at similar metallicities to those from \citet{Ryde2020}. This raises multiple questions on how these objects obtained their chemical signature and what is its astronomical origin.
 More data are needed to observationally quantify any relation between $s$-process and F nucleosynthesis in the early Galaxy.

%%%%%%%%%%%%%%%%%%%%%%%%%%%%%%%%%%%%%%%%%%%%%
\section{Conclusions}
%%%%%%%%%%%%%%%%%%%%%%%%%%%%%%%%%%%%%%%%%%%%%

We present high-resolution fluorine abundances of two CEMP stars with [Fe/H] $<-$2: HE~1429--0551 and HE~1305+0007. Our new F measurement in HE~1429--0551 makes this object the lowest metallicity star where F has been detected. Despite our small sample (current literature data is also very small: three measurements, and eight upper limits; see Section \ref{Comparison_literature}, and Figure \ref{f:F vs Fe}), our results provide valuable information on F production in AGB stars at low-metallicity and its nucleosynthesis process. We also compare our results to theoretical predictions from AGB nucleosynthesis and binary evolution model by \citet{Abate2015b} models as follow:\\

Our fluorine measurement in HE~1429--0551, A(F) = +3.93, is in reasonably agreement compared to the predicted abundance from \cite{Abate2015b}, A(F)$_{\rm Abate}$ = +4.40 (see Figures \ref{f:synthzoom} and \ref{f:Carlo_Lucatello}). The best-fitting model of \cite{Abate2015b} also reproduces the observed abundances of C and s-process elements determined by  \cite{Aoki2007}. As noted in Section \ref{target selection}, the binary status of this object is inconclusive \citep{Jorissen2016}. Without a constraint on the orbital period, \cite{Abate2015b} found that the observed abundances were best reproduced with a very long-period (+10 000 days) binary mode with an initial period of $P_i = 1.37 \times 10^5$ days, initial masses $M_{1,i} = 1.30$ \Msun and $M_{2,i} = 0.71$ \Msun. The adopted PMZ mass of the progenitor and the calculated accreted mass by HE~1429--0551 that account for the observed neutron-capture elements and F abundance is $M_{\rm{PMZ}} = 10^{-3}$ \Msun and $M_{\rm{acc}} = 0.17$ \Msun, respectively. This result strongly suggests that HE~1429--0551 is indeed a binary system which has not been monitored for a sufficiently long time to constrain its orbit.\\

Our estimated upper limit of F abundance in HE~1305+0007, A(F) $<$ +3.28, is significantly lower than the predicted value from \cite{Abate2015b}, A(F)$_{\rm Abate}$ = +4.52 (see Figures \ref{f:synthzoom} and \ref{f:Carlo_Lucatello}). \cite{Lucatello2011} also found a F upper limit in HE~1305+0007 of A(F)$_{\rm Lucatello}$ $<$ +3.5, comparable to ours if we consider the differences in the analysis (see Section \ref{Comparison_literature}).\\

All the observed upper limits in CEMP-s/r stars from \citet{Lucatello2011} and this work show low F abundances compared to CEMP-s (See figure \ref{f:Carlo_Lucatello} and \ref{f:Ba-F}). These observational results suggest that the CEMP-s/r progenitors might provide lower F abundances. The discrepancies between the observed upper limits and the predicted fluorine abundances by \cite{Abate2015b} can be due to the CEMP-s/r category type, and the theoretical approach using just $s$-process nucleosynthesis. \cite{Hampel2016,Hampel2019} used $i$-process nucleosynthesis models getting a better agreement with the chemical pattern observed in CEMP-s/r stars, including HE~1305+0007. Although the details on how the operation of the $i$-process affects the production of F has not yet been fully explored,  F abundances can be the key to understand the astronomical site of $i$-process.\\

We also show in Figure \ref{f:Ba-F} the A(F) abundances from \cite{Lucatello2011} and this work as a function of A(Ba) abundances from literature data. Although caution has to be taken in the interpretation of Figure \ref{f:Ba-F}, the data may suggest a positive correlation on Ba and F abundances in CEMP-s stars. Additionally, we note that the only three F measurements from CEMP-s objects show higher F content than those in CEMP-s/r. This could indicate differences in the relation between $s$-process and F nucleosynthesis in CEMP-s and CEMP-s/r progenitors.\\

The fluorine observations presented in this paper, alongside literature data, suggest further tests and constraints on nucleosynthesis models for CEMP star progenitors. Moreover, the relation between the production of fluorine and $s$-process elements in metal-poor AGB stars is still uncertain, and thus the true nature of its production mechanism remain unclear at these metallicities. Despite the difficulty of the analysis, it is very important to extend the study to additional CEMP stars, in order to gain new insights upon nucleosynthesis and chemical enrichment in the early galaxy.

%%%%%%%%%%%%%%%%%%%%%%%%%%%%%%%%%%%%%%%%%%%%%
\section*{Acknowledgements}
%%%%%%%%%%%%%%%%%%%%%%%%%%%%%%%%%%%%%%%%%%%%%

This work used the Immersion Grating Infrared Spectrometer (IGRINS) that was developed under a collaboration between the University of Texas at Austin and the Korea Astronomy and Space Science Institute (KASI) with the financial support of the US National Science Foundation under grants AST-1229522 and AST-1702267, of the McDonald Observatory of the University of Texas at Austin, and of the Korean GMT Project of KASI. These results made use of the Lowell Discovery Telescope at Lowell Observatory, supported by Discovery Communications, Inc., Boston University, the University of Maryland, the University of Toledo and Northern Arizona University. Parts of this research were supported by the Australian Research Council Centre of Excellence for All Sky Astrophysics in 3 Dimensions (ASTRO 3D), through project number CE170100013. We thank the referee for her/his helpful comments that improved the clarity of this paper. CA thanks Richard Stancliffe for many inspiring discussions. CK acknowledges funding from the United Kingdom Science and Technology Facility Council through grant ST/M000958/1 and ST/R000905/1. A.MG. acknowledges the support by CONICYT (Chile) through Programa Nacional de Becas de Doctorado 2014 (CONICYT-PCHA/Doctorado Nacional/2017- 72180413). A.MG. thanks Thomas Nordlander and Melanie Hampel for helpful discussions, and also dedicates this paper to all those who have been fighting for greater equality and better education in Chile.

%%%%%%%%%%%%%%%%%%%%%%%%%%%%%%%%%%%%%%%%%%%%%
\section*{Data Availability}
%%%%%%%%%%%%%%%%%%%%%%%%%%%%%%%%%%%%%%%%%%%%%

Data available on request. The data underlying this article will be shared on reasonable request to the corresponding author.



\bibliographystyle{mnras}
\bibliography{AM-G.bib} % if your bibtex file is called example.bib

%%%%%%%%%%%%%%%%%%%%%%%%%%%%%%%%%%%%%%%%%%%%%%%%%%

% Don't change these lines
\bsp	% typesetting comment
\label{lastpage}
\end{document}